\documentclass{article}

\usepackage{PRIMEarxiv}

\usepackage[utf8]{inputenc} 
\usepackage[T1]{fontenc}    
\usepackage{hyperref}       
\usepackage{url}            
\usepackage{booktabs}       
\usepackage{amsfonts}       
\usepackage{nicefrac}       
\usepackage{microtype}      
\usepackage{lipsum}
\usepackage{graphicx}
\usepackage{amsmath}
\graphicspath{{media/}}     

\title{Exploring How Fair Model Representations Relate to Fair Recommendations}

\author{
  Bj{\o}rnar Vass{\o}y, Benjamin Kille, Helge Langseth\\
  Norwegian University of Science and Technology \\
  Trondheim, Norway \\
  \texttt{bjonar.vassoy@ntnu.no} \\
}

\begin{document}
\maketitle


 
\begin{abstract}
  One of the many fairness definitions pursued in recent recommender system research targets mitigating demographic information encoded in model representations. Models optimized for this definition are typically evaluated on how well demographic attributes can be classified given model representations, with the (implicit) assumption that this measure accurately reflects \textit{recommendation parity}, i.e., how similar recommendations given to different users are. We challenge this assumption by comparing the amount of demographic information encoded in representations with various measures of how the recommendations differ. We propose two new approaches for measuring how well demographic information can be classified given ranked recommendations. Our results from extensive testing of multiple models on one real and multiple synthetically generated datasets indicate that optimizing for fair representations positively affects recommendation parity, but also that evaluation at the representation level is not a good proxy for measuring this effect when comparing models. We also provide extensive insight into how recommendation-level fairness metrics behave for various models by evaluating their performances on numerous generated datasets with different properties.
\end{abstract}




\maketitle

\section{Introduction}
Fairness does not have a definition shared throughout the academic community. Different definitions can often be exemplified in whether one believes the data to be accurate or biased. For instance, in a career recommender dataset, one may argue that the data is biased and that different demographic groups have more similar preferences than the data indicates, e.g., many historically gendered occupations are more balanced today. In other domains where demographic preference differences have remained more static, it may be more fitting to trust the data and rather make sure that the recommendation system performs equally well for each group. This paper focuses on a definition that falls under the former example, \textit{neutral representation} \cite{consumer_fair}, and from here onwards any use of \textit{fair} or \textit{fairness} will refer to this definition.

Neutral representation is a fairness objective that focuses on mitigating the encoding of user demographics in model representations. This fairness definition focuses on users rather than items/providers, and would be categorized as consumer-side fairness (C-fairness) in the taxonomy proposed in \cite{multi_sided_fair}. Its motivation is that if a model representation does not correlate with the user's demographics, the demographics do not influence the model's treatment of said user. Models optimized for providing users \textit{recommendations} that do not correlate with their demographics fall under methods optimized for recommendation parity in the consumer-side fairness taxonomy proposed in \cite{consumer_fair}. While perfect optmization of neutral representation fairness theoretically should lead to perfect recommendation parity, we are unaware of any extensive research into how the two fairness objectives and their associated metrics relate for different models, levels of optmization, and different dataset properties. Further, while there are well-established metrics for measuring how different recommendation sets are for rating-prediction recommendation \cite{beyond,fair_metric_ashokan,kamishima_enhancement_2012,kamishima_model-based_2016} where the models predict how each user will rate every item, it is considerably harder to measure similar properties for models that output ranked top-k recommendations. A metric for comparing such sets of recommendations must decide how to handle ranking, disjoint sets of items, and potential aggregation of items for user groups, and it should ideally be interpretable. A few metrics have been proposed, but, unlike the evaluation of neutral representation, there are no go-to metrics for evaluating recommendation parity of top-k recommendations. Our work explores how various models perform in both the standard neutral representation metric and recommendation parity metrics for different hyperparameters on different datasets. We also look into classifying users' demographics given their recommendations, similar to how neutral representations are evaluated. In particular, we are interested in gaining insights into the following questions:

\begin{itemize}
    \item Are neutrality estimates of model representations good proxies for how much user demographics influence recommendations?
    \item Are neural classifiers apt for estimating how much user demographics influence top-k recommendations?
    \item How do metrics for comparing sets of ranked recommendations behave for different models and dataset properties?
\end{itemize}

\section{Background}
Models optimized for neutral representation are usually evaluated by estimating how well classifiers can be trained to identify demographic classes given model Representations. The success of said classifiers is often measured by Area Under the ROC-Curve (AUC), which is defined as the area under the curve defined by the True Positive Rate and False Positive Rate for each classification threshold. AUC is preferred over metrics such as accuracy since it also accounts for False Negatives, and considers all threshold values without requiring the specification of one for each trained model. Successful classifiers have thresholds for which True Positive Rate > False Positive Rate, which results in AUC > 0.5, with 1.0 being optimal. For neutral representation, we want the classifier to fail since this implies that the representations no longer encode demographics, at which point its performance matches that of a random classifier with AUC = 0.5. Note that binary classifiers that achieve AUC < 0.5 can have their class probability inverted to achieve AUC > 0.5, and AUC = $x$ is usually considered equivalent to $1-x$.

Among the different types of recommender systems, we focus on models that provide users with top-k ranked lists of the items predicted to be the most relevant for each user. The data used for training our models consists of user interactions, typically interaction types that indicate the user has an interest in the item. Similar to most research into this recommendation setting, we do not allow the recommendation of items the user has already interacted with.

\section{Related Work}
Many recommender systems are optimized for neutral representation and evaluated by training classifiers to identify demographic classes based on model representations. Among the first models are \cite{privacy} and \cite{compositional}. The former proposes to directly train adversarial classifiers to predict demographics given user latent factors in a matrix factorization model and use gradients from these classifiers to update the latent factors to better fool said classifiers. The latter instead proposes to train filter networks for filtering out individual demographic features from representations, again using adversarial classifiers. This model allows users to specify which demographic features to filter out. User control is also supported by \cite{afrl} and \cite{opt_in}, which interface with pre-trained models to offer optional fairness consideration. The authors of \cite{afrl}  apply information theory and adversarial classifiers to split representations of an underlying model into one neutral and several demographic-specific representations. A second part of the approach comprises a network trained to re-combine the neutral representation and the allowed demographic representations, such that the final representation can be used in the underlying model. The authors of \cite{opt_in} instead train classifiers to identify how the input can be perturbed to obfuscate the demographic information encoded in Variational Autoencoder (VAE) latent representations, which is applied iteratively by adding and removing items. This approach also provides the user transparency into how their input has been changed to hide their demographics and how this changes their recommendations.

The methods proposed in \cite{debayes} are optimized for neutral representation without using adversarial classifiers. Specifically, the authors fit probabilistic methods with priors informative of user demographics to encourage the other parts of the models to focus on non-demographic aspects. During evaluation, they replace the informative priors with neutral ones. The authors of \cite{consumer_fair} provide a detailed overview of recommender systems optimized for neutral representation and methods evaluated by training auxiliary classifiers to classify demographics given model representations.

The work presented in \cite{metric_eval} is similar to our work in thoroughly exploring the behaviour of fairness-related metrics for different dataset properties. They focus on provider-side fairness (P-fairness) \cite{multi_sided_fair} and metrics that jointly evaluate recommendation performance and item exposure/envy. The paper, like ours, concludes with suggestions for using the considered metrics and notes on the strengths and weaknesses of the metrics based on the presented results and observations.

The related works we have highlighted either propose models optimized for neutral representation fairness, or explore fairness evaluation in depth. Our work partly combines the two by thoroughly exploring how such models and their optimization relate to fairness evaluation of output rather than model representation.

\section{Method}
We propose two methods for estimating how much demographics have influenced recommendations and apply existing methods for generating synthetic datasets to explore how the models and metrics perform for different data properties.

\subsection{Recommendation AUC}
As an alternative to measuring how much the model representations reveal about a user's demographic features, we want to measure the same given ranked sequences of recommendations. This is a challenging task given how sparse recommendation data is, the (often) large number of items, and how item popularity typically follows a Power-Law distribution \cite{power-law}. Most items will only account for a few entries in the dataset and will only be observed along a small set of the total number of items in the recommendation lists in which they appear. This causes a lot of uncertainty for data-driven models, as it is difficult to know how rare these relations are.

\subsubsection{Neural AUC}
A neural classifier is prone to learn spurious correlations of rare items. To mitigate this issue, we first train item embeddings using a Skip-Gram approach. Specifically, for each item liked by a user, we sample up to five \textit{neighbour} items among other items liked by the user, and train a neural embedding network to maximize the predicted probability of the neighbour items given the input item. Unlike in other embedding settings, our training data does not have sequence information, so neighbours are sampled among all items liked by the same user. The embedding layer of this network is then extracted for the recommendation classifiers. This mitigates overfitting by having the embedding layer constrained by what the embedding model encoded, and limits the recommendation classifiers' abilities to push rare and demographically polarized items to extreme points in the embedding space. In other words, the recommendation classifier has to work with the more general learned features to separate the cases it is presented with.

Multiple settings and architectures were tested for the sequence model: whether to use pre-trained embeddings, feedforward- vs RNN- vs Transformer-encoder-based layers, and whether to add information on how demographically polarized each item is in the training data. Our tests showed that generalization improved when enabling pre-trained embeddings using the described Skip-gram approach. RNN and Transformer-encoder-based models outperformed the feedforward model, but we observed negligible difference between the two. Preliminary testing did not reveal noticeable improvement from including information on how demographically polarized each item is, but we opted for keeping it based on how well this information is used in the alternative Recommendation AUC measure.

\paragraph{Note on different training/evaluation strategies} The Representation AUC evaluated in the VAE-based models we compare was measured using Logistic regression models trained exclusively on testing data and evaluated using cross entropy. The recommendation classifiers used to measure Neural AUC are trained on the training data and only evaluated on the testing data. The former may be more accurate in scenarios where training representations and testing representations follow different distributions, i.e., poor generalization, but may suffer from high variance in smaller datasets since the validation part is small. The latter has more training data, but will be inaccurate if the model is overfitted. We concluded that this training/evaluation difference was warranted for the following reasons: We verified that both approaches yield insignificant differences in expected values for both types of AUC. Further, since the recommendation classifiers require way more computation to train, it would be very demanding to train them enough times to counteract the high variance of the former strategy.

\subsubsection{Demographic Ratio AUC}
As a simple alternative to the Neural network-based Recommendation AUC, we use the median \textit{demographic ratio} of the items in the recommendation list. An item's demographic ratio is defined as the number of times the minority user group has interacted with the item, divided by the total number of interactions with said item. For items preferred by one user group, the resulting ratio differs from the ratio of minority users in the dataset. We use the median demographic ratio of the items recommended to a user as a predictor for their demographics and calculate AUC as usual.
\begin{equation}\label{eq:demratio}
    \text{DemographicRatio}_i = \frac{\sum 1\{d_{u,i}\in \mathcal{D} \land u \in U_{\text{minority}}\}}{\sum 1\{d_{u',i}\in \mathcal{D} \land u' \in U\}}
\end{equation}
Here, $d_{u,i}$ indicates that user $u$ has interacted with item $i$, $\mathcal{D}$ is the set of all interactions, $U$ is the set of users, and $U_{\text{minority}}$ is the set of users belonging to the smallest demographic group.

\subsection{Dataset Generation}
The synthetic datasets were generated by the approach proposed in \cite{coppolillo2024genrecflexibledatagenerator}. The generation process can be summarized by the following equations:
\begin{align}
    \text{pop}_i &\sim \text{LongTail}(\lambda)\\
    \hat{d}_{u,i} &\sim \text{Bernoulli}(t_{u,i}^{\delta (1-\text{pdf}(\text{pop}_i))}) 
\end{align}
where $\text{LongTail}$ can be any long-tail distribution (Power-Law, Stretched Exponential, Log-Normal etc) with parameters $\lambda$ and density function $\text{pdf}$. Further, $\text{pop}_i$ is item $i$'s popularity score, $\delta$ is a control parameter for item popularity and $\hat{d}_{u,i}$ indicates that item $i$ is a candidate item of user $u$. The variable $t_{u,i}$ represents the underlying utility of item $i$ for user $u$ and is defined as a factorization of latent Dirichlet variables. The dimensions of the latent variables represent item features and allow the specification of item and user categories. The different categories are related to features through Dirichlet priors, while features not associated with the category are given prior parameters of $\epsilon$, which serves as a hyperparameter.

A user's data is sampled from the candidate items $\hat{d}_{u,i}$ since the Bernoulli sampling of the candidate items will typically lead to the \textit{number} of candidates per user being normally distributed. The second sampling instead enforces this number to follow a LongTail distribution as in many real dataset. The sampling is done as follows:
\begin{align}
    n_u' &\sim \text{LongTail}(\beta)\nonumber\\
    n_u &= n_u' + \tau\\
    \mathcal{D}_u &= \{i_j|i_j \sim \text{Uniform}(\mathcal{\hat{D}}_u),j=1,...,n_u\}
\end{align}
where $n_u$ is the number of interactions for user $u$, $\tau$ is a minimum number of interactions and $\mathcal{D}_u$ is final set of item interaction of user $u$ in the synthetic dataset.

We used Log-Normals for both $\text{LongTail}$ distributions and found realistic values for the parameters $\lambda$ and $\beta$ by fitting Log-Normals to the number of interactions per item and the number of interactions per user in the Movielens 1 Million dataset \cite{movielens}. 



\section{Experimental Setting}
Our experiments are run on the Movielens 1 Million dataset \cite{movielens} and synthetically generated datasets. The Movielens dataset contains roughly 6k users, 3.5k items, and we extract binary demographic attributes for Gender and Age for all users. The minority Gender group makes up 28.3\% of the users, while the same number is 23.6\% for Age. The training-test split was performed such that each user is only represented in one dataset, as the considered models are designed to recommend any user when given their user data. We selected $k=40$ for the number of recommendations considered in each metric since we hypothesized that typical values like 10 would reflect very little demographic influence for fairness optimized methods. Later testing of this hypothesis revealed smaller differences than anticipated, but we opted to keep $k=40$ because it yielded results with smaller variance. Our code is publicly available\footnote{https://github.com/BjornarVass/fair\_rec\_eval}.

\subsection{Models}\label{sec:models}
We utilized existing implementations from \cite{opt_in} for all VAE-based models. For implementation details, please refer to this paper and its accompanying code. We performed hyperparameter searches to determine the optimal hyperparameters for the models in the Movielens dataset, as well as to identify viable parameters when varying the degree of fairness weighting in specific tests. For experiments using synthetic datasets, we only considered models without hyperparameters or models that are insensitive to hyperparameter choices.

\subsubsection{\texttt{POP}}
This model is the popularity baseline. Each user is recommended the $k$ most popular items among the items the user has not already interacted with. \texttt{POP} is fair since it applies the same heuristic for each user regardless of demographics.

\subsubsection{\texttt{Rand}}
This model is the random baseline. Each user is given a random list of recommendations among the items the user has not liked. \texttt{RAND}'s heuristic is fair.

\subsubsection{\texttt{Dem. POP}}
This model is a popularity baseline on a demographic level. Users of a demographic group are given recommendations based on the most popular items among their demographic group. \texttt{Dem. POP} is designed to be discriminatory.

\subsubsection{\texttt{Max Division}}
This model is a baseline that recommends the most demographically divisive items to the demographic groups that prefer them. Items that occur fewer than five times in the training set are excluded. \texttt{Max Division} is designed to be discriminatory.

\subsubsection{\texttt{VAE}}
The \texttt{VAE} \cite{fairvae} model is a popular VAE-based recommender system. Trained as a Noisy-VAE to reconstruct lists of items interacted with by the user. Recommendations are extracted from the noisy reconstruction based on probabilities assigned to items the user has not liked. \texttt{VAE} is not optimized for fairness.

\subsubsection{\texttt{VAERel} and \texttt{VAE2Adv}}
The \texttt{VAERel} and \texttt{VAE2Adv} models \cite{opt_in} are both fair extensions of \texttt{VAE} that iteratively perturb the user input to obfuscate demographic information. Perturbations are identified by training classifiers to classify demographics given the VAE latent representation and propagating neural gradients to the inputs where they inform which additions/removals would best obfuscate. Both methods are given a budget that indicates the maximum number of items that can be perturbed, but \texttt{VAE2Adv} differs in training a second group of classifiers to take over after, at most, half of the budget has been spent. \texttt{VAE2Adv} typically improves fairness further since the first group of adversarials may focus on items that overshadow other items through strong correlations, while the second group can learn weaker correlations after the former has been dealt with. Both are trivial to tune for any dataset properties.

\subsubsection{\texttt{VAEAfrl*}}
The \texttt{VAEAfrl*} model is another fair extension of \texttt{VAE} based on \cite{afrl}. The model alternates between training adversarial classifiers to classify demographics given a filtered VAE latent state and the filtering network to fool the adversarials, i.e., generative adversarial networks (GANs). A second part is trained to combine the filtered representation with (extracted) demographic information based on the user's permission. The combined representation is used with the underlying recommender system. Further details on how the approach proposed in \cite{afrl} was adapted for VAE and a different evaluation setting are found in \cite{opt_in}.

\subsection{Metrics}
\subsubsection{Item ratio}
Item ratio is proposed in \cite{opt_in} as a simple, intuitive metric for measuring how far the recommendations are from the ideal case where each item is equally likely to be recommended for each demographic group. The metric relies on measures for each items similar to Demographic Ratio (Eq \ref{eq:demratio}), only differing in counting item occurrences in the training data rather than the recommendations. For each item, it calculates the absolute difference of this measure and the ratio of minority users in the dataset, and aggregates the items by mean. Each item will contribute 0 in a fair scenario since the ratios match. The main drawback of this metric is related to fraction precision, i.e., the best possible ratio of an item with few recommendations may be far from the minority ratio. This issue is exacerbated for minority ratios that require large numbers of observations to approximate accurately.

\subsubsection{Kendall-Tau}
Kendall-Tau is an established metric for comparing lists containing the same set of items. We consider an extension\footnote{https://xebia.com/blog/using-kendalls-tau-to-compare-recommendations/} of the metric for the recommendation setting, where top-k recommendations often contain different sets of items. The extended metric is 1 for identical lists, and -1 for lists with disjoint sets of items. We aggregate recommendations for each demographic group by counting the occurrences of each item in the users' top-k recommendations and ranking the items by occurrences. The top-k items among these aggregations are then compared.

\section{Results}
The default synthetic data is generated with 4000 users, 4000 items, and the minority demographic group accounts for 30\% of the users. We based these settings on the movielens dataset, and opted for numbers well suited for mental arithmetics. Each experiment specifies any changes to these values in text or plotted axes. The VAE-based models are all trained at least 5 times for each set of hyperparameters, and we use early-stopping to train the models for varying dataset properties and settings. Tests that vary dataset hyperparameters generate at least five datasets for each set of parameters.

\subsection{Comparing Representation and Recommendation AUC}
We explore two approaches for comparing Recommendation AUC and Representation AUC: The first approach keeps model hyperparameters static while generating synthetic datasets for different $\epsilon$ settings, i.e., varying how much the preferences of the two demographic groups overlap/diverge. In the second approach, the dataset is kept static and we train multiple versions of each model while varying hyperparameters controlling the importance of the fairness objectives.

\begin{figure}
    \centering
    \includegraphics[width=1.0\linewidth]{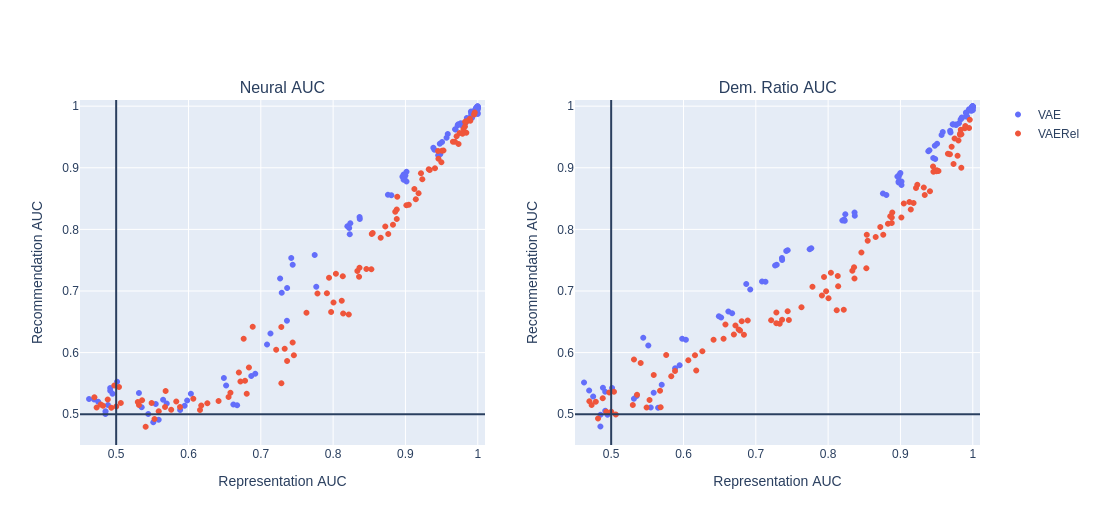}
    \caption{Recommendation AUC plotted against Representation AUC for synthetic datasets with different $\epsilon$ parameters. \texttt{VAE} and \texttt{VAERel}.}
    \label{fig:auc_synth1}
\end{figure}
\begin{figure}
    \centering
    \includegraphics[width=1.0\linewidth]{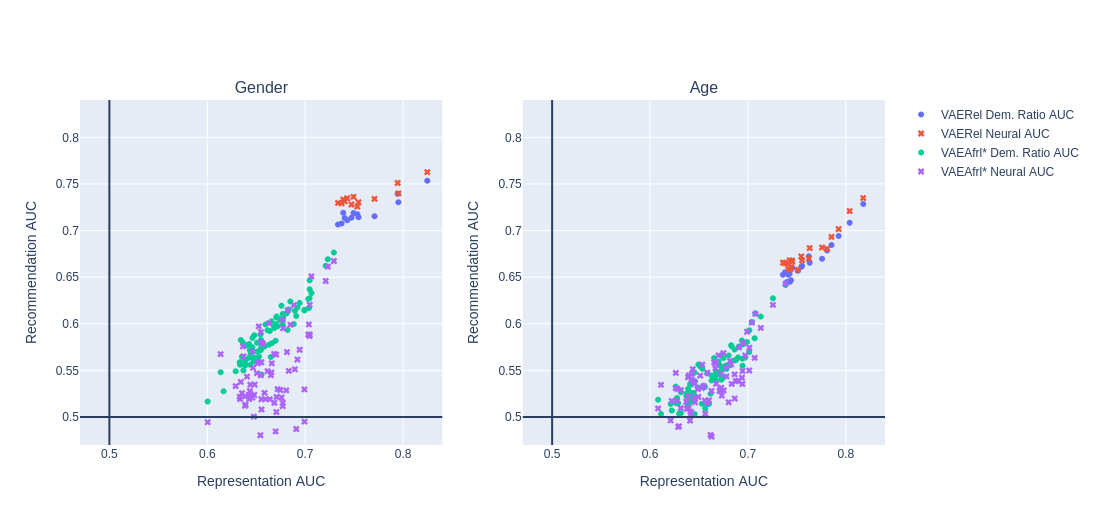}
    \caption{Gender and Age Recommendation AUC plotted against Representation AUC for the Movielens 1M datasets with different model parameters. \texttt{VAERel} and \texttt{VAEAfrl*}.}
    \label{fig:auc_real1}
\end{figure}
\begin{figure}
    \centering
    \includegraphics[width=1.0\linewidth]{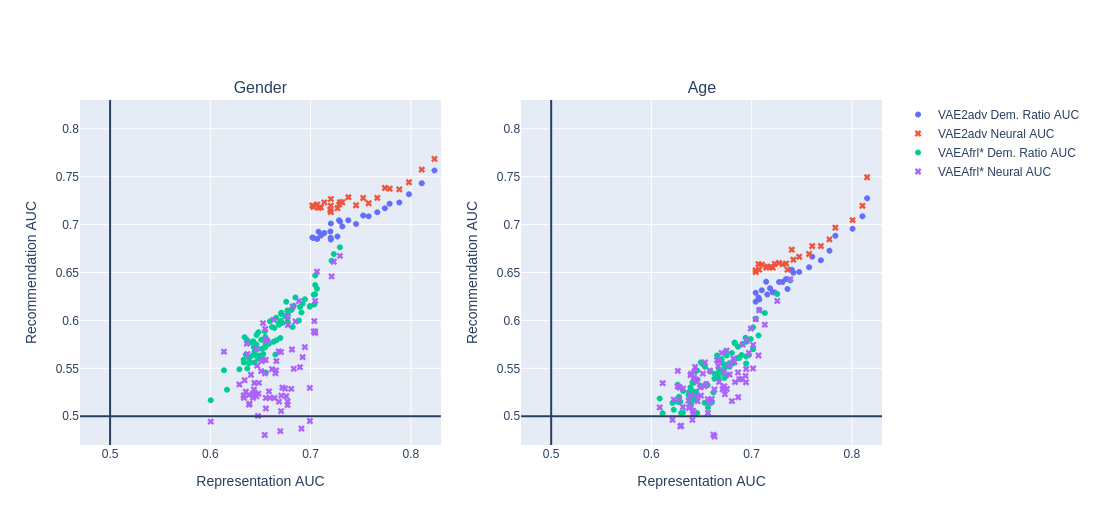}
    \caption{Gender and Age Recommendation AUC plotted against Representation AUC for the Movielens 1M datasets with different model parameters. \texttt{VAE2adv} and \texttt{VAEAfrl*}.}
    \label{fig:auc_real2}
\end{figure}
\begin{figure}
    \centering
    \includegraphics[width=1.0\linewidth]{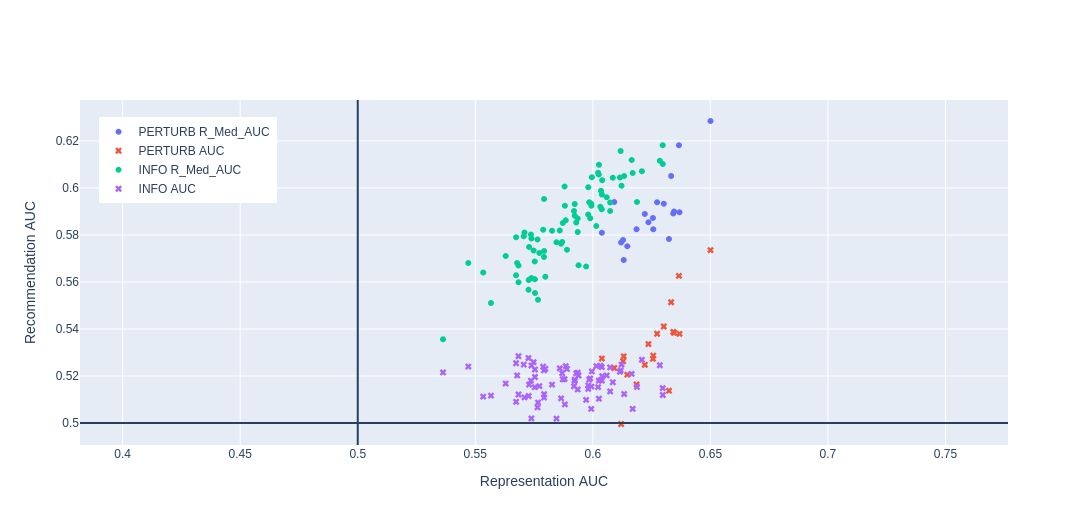}
    \caption{Recommendation AUC plotted against Representation AUC in a synthetic dataset with $\epsilon=0.74$ parameters.}
    \label{fig:auc_synth2}
\end{figure}
\begin{figure}
    \centering
    \includegraphics[width=1.0\linewidth]{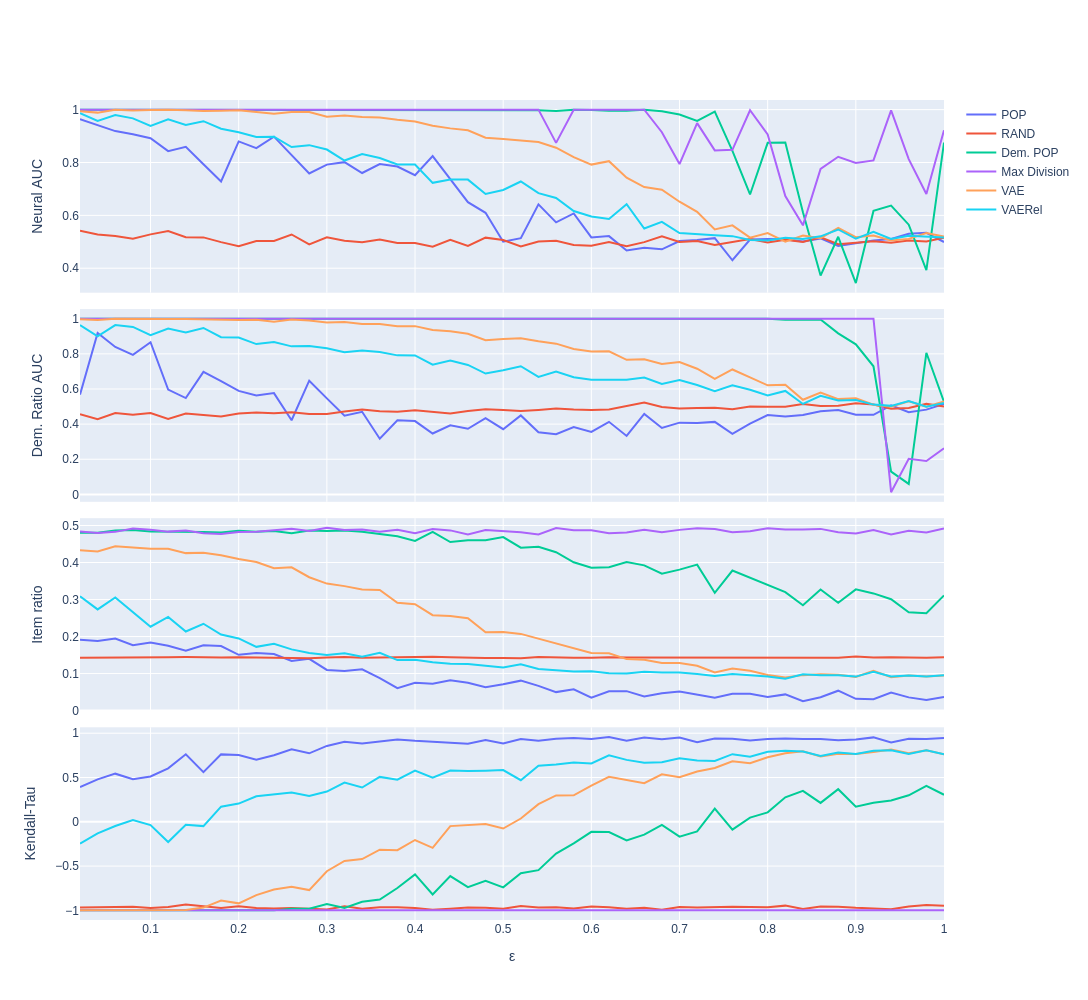}
    \caption{Recommendation metrics plotted for different models and dataset $\epsilon$-parameters.}
    \label{fig:eps}
\end{figure}
\begin{figure}
    \centering
    \includegraphics[width=1.0\linewidth]{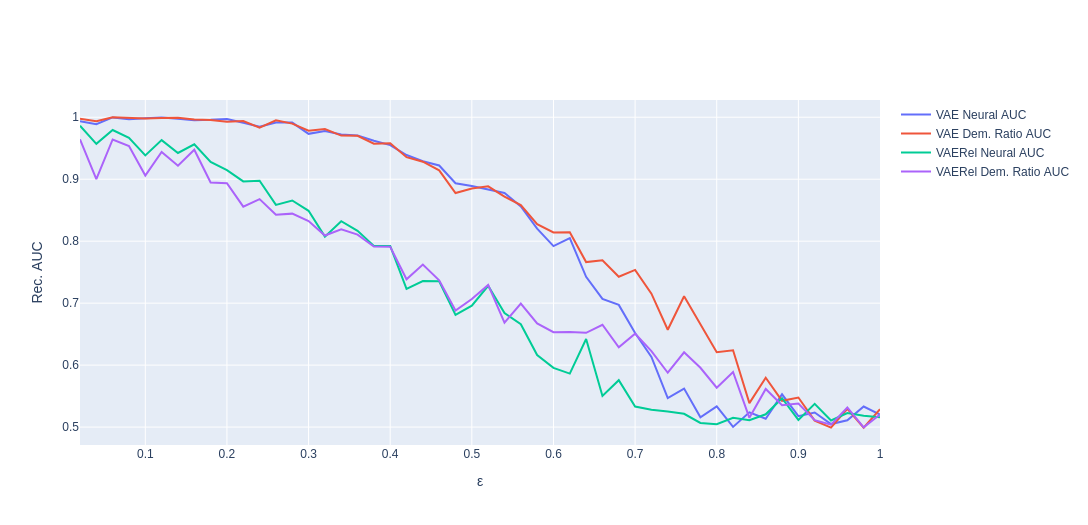}
    \caption{Comparison of \texttt{VAE} and\texttt{VAERel} Demographic Ratio AUC plotted for each dataset $\epsilon$-parameters.}
    \label{fig:eps_comp}
\end{figure}
\begin{figure}
    \centering
    \includegraphics[width=1.0\linewidth]{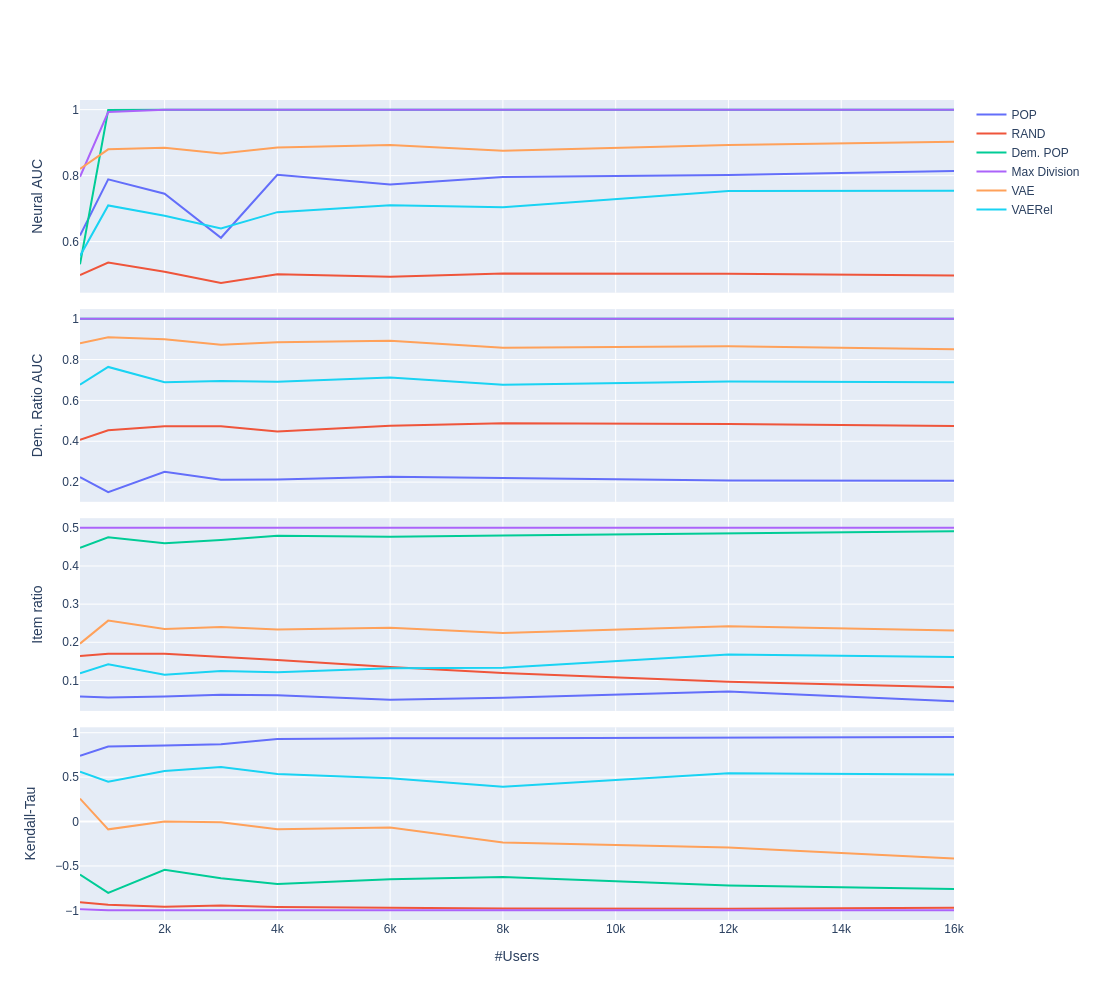}
    \caption{Recommendation metrics plotted for different models and number of users in dataset.}
    \label{fig:nu}
\end{figure}
\begin{figure}
    \centering
    \includegraphics[width=1.0\linewidth]{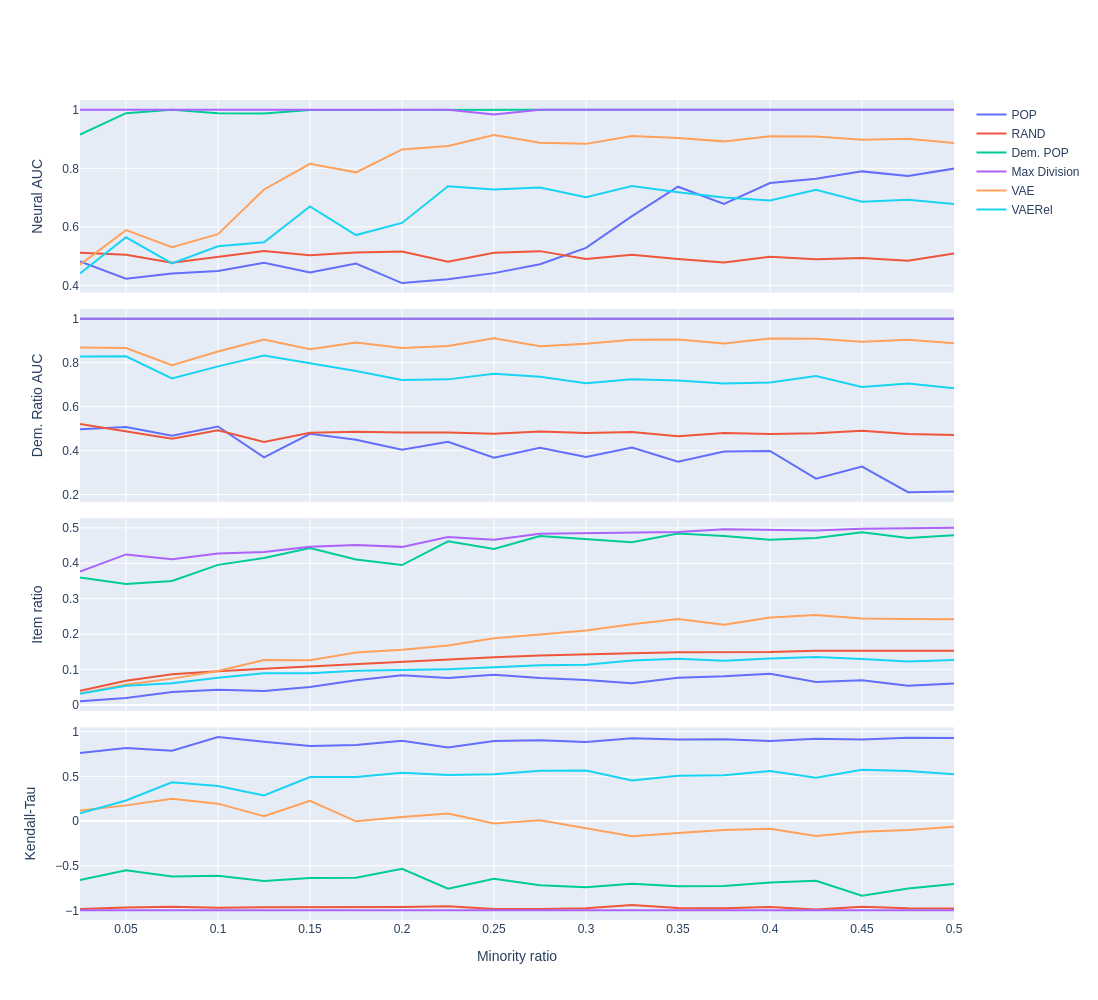}
    \caption{Recommendation metrics plotted for different models and dataset minority ratio.}
    \label{fig:skew}
\end{figure}

\subsubsection{Static Models, Different Datasets}
We generated 50 datasets for each $\epsilon$ value between 0.02 and 1.00 with 0.02-sized intervals. Figure \ref{fig:auc_synth1} reveal three key insights:
\begin{enumerate}
    \item For Representation AUC higher than 0.65, \texttt{VAE} and \texttt{VAERel} indicate a roughly linear relation between Neural and Representation AUC. For lower Representation AUC, the slopes flatten, and the Neural classifier fails to pick up noticeable differences between the recommendation lists of different demographic groups.
    \item For both Recommendation AUC metrics, the plots reveal that the fair model \texttt{VAERel} can achieve lower Recommendation AUC over a substantial Representation AUC interval. This implies that the fair models and their active minimization of Representation AUC are more efficient at reducing \textit{Recommendation} AUC than the passive removal of information by increasing dataset $\epsilon$. Reducing Representation AUC by 0.1 through switching from \texttt{VAE} to \texttt{VAERel} results in a greater reduction of Recommendation AUC than the same reduction of Representation AUC through increasing $\epsilon$ without switching the model.
    \item The Demographic Ratio AUC of \texttt{VAE} has an almost perfectly linear relationship with Representation AUC, i.e., the non-fair baseline will encode and output most of the demographic differences in the generated dataset.
\end{enumerate}


\subsubsection{Different Models, Static Dataset}
In the Movielens dataset, we consider two demographic features: Gender and Age (Figure \ref{fig:auc_real1}). The two fair models \texttt{VAERel} and \texttt{VAEAfrl*} are combined in the same plot to cover more of the AUC ranges and to highlight potential differences in how their optimization performs for different types of AUC.
\begin{itemize}
    \item For both Gender and Age, Recommendation AUC trend towards 0.5 for Representation AUC around 0.6, i.e., recommendations may become fair while the representations still reveal demographics.
    \item Neural AUC varies more and is typically lower than Demographic Ratio AUC for smaller Representation AUC values, in particular for Gender.
    \item The relation between Neural AUC and Demographic Ratio AUC is flipped for large Representation AUC and different models. This seems to be both model and dataset-dependent (Figures \ref{fig:auc_real2}, \ref{fig:auc_synth2} and \ref{fig:eps_comp}).
\end{itemize}

If \texttt{VAERel} is replaced with the more fair \texttt{VAE2adv}, as seen in Figure \ref{fig:auc_real2}, it becomes clear that the two different fairness approaches do not have overlapping curves for the same Representation AUC. In other words, different approaches for minimizing Representation AUC can have different impacts on Recommendation AUC.




A final comparison of Representation and Recommendation AUC when training \texttt{VAERel} and \texttt{VAEAfrl*} with different hyperparameters is shown in Figure \ref{fig:auc_synth2}. Here, we look at synthetic datasets with $\epsilon=0.74$ to see how the same comparison looks when the demographic preferences differ substantially less than in the Movielens dataset. Key differences are that Neural AUC is lower than Demographic Ratio AUC for the full Representation AUC interval and both models. While the variance is greater and Representation AUC ranges are mostly disjoint, the plot indicates that \texttt{VAERel} achieve similar Recommendation AUC as \texttt{VAEAfrl*} but at $\approx0.04$ worse Representation AUC. This is the opposite of the Movielens results, i.e., one method appears more efficient than the other in a dataset with large demographic differences, but less efficient in datasets with smaller demographic differences. For these datasets, none of the models get close to 0.5 Demographic Ratio AUC, and the slopes are flatter than for Movielens


\subsection{Model Behaviour for Different Datasets}
In this section, we compare models for different metrics while varying dataset parameters. We explore the effect of varying how different the demographic groups are, the number of users in the dataset, and how skewed the demographic groups are. Unlike previous sections, we also compare the VAE-based methods with other baselines. Unfortunately, \texttt{VAEAfrl*} proved difficult to tune for these experiments as the GAN setup is highly sensitive to dataset properties and hyperparameters. The only fair VAE-based model considered in this Section is \texttt{VAERel}.

\subsubsection{Varying $\epsilon$}\label{sec:vary eps}
The hyperparameter $\epsilon$ functionally defines how similar the preferences of different demographic groups are, where $\epsilon=0$ results in disjoint preferences and $\epsilon=1$ in overlapping preferences. The results are shown in Figure \ref{fig:eps}, and we structure observations by model.

\paragraph{\texttt{POP}} The popularity baseline is a bit more unstable than other models since it is heavily impacted by how the most popular items are distributed over items preferred by different demographic groups, especially because the groups have skewed sizes. For small $\epsilon$, the different groups like no or few overlapping items, the majority class tend to dominate the popular items, and both types of Recommendation AUC are high. Interestingly, mid-valued $\epsilon$ results in Demographic Ratio AUCs lower than 0.5. Here, the preferences have started to overlap, and popular items skewed towards one class may appear more often in recommendations given to the other class. This is because many of the users in the group that prefer the item have already interacted with it and will not be recommended the item. So, the fact that we do not recommend items to users who have already interacted with an item results in the Popularity baseline outputting different distributions of recommendations for different demographic groups, despite the general heuristic being fair. Neural AUC appears to capture the pattern and maximize the output probabilities of the correct class, thus, AUC is greater than 0.5. The model's Item ratio and Kendall-Tau performance get better for larger $\epsilon$ and is also the best-performing model for these metrics and datasets. The great number of recommendations of the same items minimizes the precision issue of Item ratio, and the fair heuristic results in similar items in the top recommendations.

\paragraph{\texttt{RAND}} The random baseline performs very consistently for different metrics and different $\epsilon$ values. Neural and Demographic Ratio AUC indicate that this model also produces slightly different recommendations for small $\epsilon$, i.e., the recommended lists will differ somewhat since the items the users have interacted with are excluded from the random sample. The random sample consistently results in aggregated group recommendations with disjoint sets of items, giving the worst Kendall-Tau value of -1. Finally, this model is expected to recommend an item equally to all users, but by distributing the recommendations to all items, each item will only appear a few times. The former is good for the Item ratio metric while the latter can decrease performance, resulting in \texttt{RAND} being outperformed by \texttt{POP} for most $\epsilon$  values and by both VAE-based models for larger values. 

\paragraph{\texttt{Dem. POP} and \texttt{Max Division}} The two models designed to be discriminatory share many similarities. Both perfectly reflect the users' demographics for all but the largest $\epsilon$, but are very unstable for datasets where preferences mostly overlap. The latter is because the selection of popular items among each group, and the items that are the most divisive, becomes arbitrary once preferences mostly overlap. While \texttt{Dem. POP} achieves a Kendall-Tau of greater than 0 and improved Item ratio for $\epsilon$ close to 1, \texttt{Max Division} achieves the worst possible Kendall-Tau and worst observed Item ratio across all $\epsilon$. 

\paragraph{\texttt{VAE} and \texttt{VAERel}} The VAE models are best discussed together as the latter is a fair extension of the former. As expected, \texttt{VAERel} outperforms \texttt{VAE} on all metrics, but their performance mostly overlap for very large $\epsilon$. As seen in other experiments, Neural AUC is reduced to the perfect 0.5 for datasets where the preferences of the demographic groups overlap. This is also the case for Demographic Ratio AUC at very large $\epsilon$. Finally, when comparing the two types of Recommendation AUC of both VAE-based models in Figure \ref{fig:eps_comp}, we see that both models have a higher Demographic Ratio AUC for larger $\epsilon$, but \texttt{VAERel}'s Neural AUC is higher for smaller $\epsilon$ (\texttt{VAE} achieves AUCs of 1 for most of this area so they cannot be compared). This corroborates results in Figure \ref{fig:auc_real1}.

\subsubsection{Varying Number of Users}
Ideally, the metrics should all be very consistent when varying the number of users, as the demographic properties of the datasets do not change. Our results are found in Figure \ref{fig:nu}, and we structure the observation and discussion by metric. Most trends are shared by all models, and there are only a few surprising results for each metric.


\paragraph{Neural AUC} All models see a bump in AUC  when increasing the number of users from 500 to 1000, likely because the amount of training data limited the performance in the smallest datasets. \texttt{POP} has a surprising drop for 3000 users, and is hypothesised to be related to the previously noted instability of this model, coupled with the relatively small number of users. Finally, \texttt{VAERel} appears to achieve a slightly worse AUC score on datasets with more users which is also seen in its Item ratio performance. This is perhaps related to the model relying on adversarial classifiers to introduce fairness, and the increased number of users impacts their training. However, similar trends are not seen for Demographic Ratio AUC and Kendall-Tau.


\paragraph{Demographic Ratio AUC} Most models are fairly consistent here, with minor fluctuations for smaller numbers of users. As observed and discussed in Section \ref{sec:vary eps}, the Demographic Ratio AUC measured for \texttt{POP} is substantially below 0.5.

\paragraph{Item ratio} It is apparent that increasing the number of users improves the Item ratio achieved by \texttt{RAND}. This is not surprising given the discussed precision issue related to this metric, i.e., with more observations of each item, the best recommendation ratio(s) will be closer to the demographic ratio of the dataset and incur smaller penalties.

\paragraph{Kendall-Tau} The Kendall-Tau metric is also fairly consistent for all models, but it is worth noting that \texttt{VAERel} roughly fluctuates around the same value while \texttt{VAE} is slowly deteriorating.

\subsubsection{Varying Demographic Split}
The final suite of tests shown in Figure \ref{fig:skew} compares metrics while varying the relative size of the smallest demographic group (minority ratio).

\paragraph{\texttt{POP}} The popularity baseline displays some unexpected behaviour for Neural AUC. As for other models, the neural model struggles to learn anything for very skewed demographic groups. However, the popularity baselines achieve well below 0.5 AUC for smaller minority ratios before jumping to levels more comparable with the VAE-based models. The leading hypothesis is that the neural models struggle to deal with the long-tail popularity distribution and how the few most popular items are distributed among items preferred by the different groups. Instability/fluctuation is also clear in its Demographic Ratio AUC, but to a lesser degree, and shows a clear deteriorating trend. As seen in other plots, \texttt{POP}'s Demographic Ratio AUC stays under 0.5 and worsens by moving further away from 0.5. Finally, despite these trends and fluctuations, Kendall-Tau remains largely consistent for all but the smallest minority ratios, and the Item ratio follows similar curves as the other models.

\paragraph{\texttt{RAND}} Consistent across all metrics.

\paragraph{\texttt{Dem. POP} and \texttt{Max Division}} Consistently performs the worst for all metrics. \texttt{Max Division} either overlaps with \texttt{Dem. POP} or performs slightly worse.

\paragraph{\texttt{VAE} and \texttt{VAERel}} \texttt{VAERel} performs better than \texttt{VAE} for all minority ratios on all metrics. Both models indicate that the models used for estimating Neural AUC struggle for minority ratios smaller than 0.225 as the plots show substantial fluctuation that starts at AUC of about 0.5. At greater ratios, both curves appear mostly flat. Their Demographic Ratio AUC starts off at roughly the same value for low minority ratios, but settles into more stable values at roughly the same ratio threshold, with \texttt{VAE} worsening and \texttt{VAERel} improving. This is very similar to their behaviour for both Item ratio and Kendall-Tau, with the caveat that all models worsen in Item Ratio as minority ratio increases, but \texttt{VAE} worsens at a noticeably higher rate than \texttt{VAERel}.
\paragraph{Note on Item ratio}
The Item ratio of all models worsens the larger the minority ratio, which is largely explained by the following observations: The smaller the minority ratio, the less the item ratios of 0 (none of the minority class were recommended the item) differ from the minority ratio. Further, fewer items preferred by the minority class will make the minimum requirement for being considered in the metric because they are not recommended to enough users, and items preferred by the majority will make up the majority of considered items.

\section{Discussion}
\subsection{Representation AUC Larger Than Recommendation AUC}
Representation AUC will always be an upper bound of Recommendation AUC, and there are many potential explanations for Recommendation AUCs being substantially smaller. We have shown that our implementation of Neural (Recommendation) AUC performs worse than the simple Demographic Ratio AUC when demographic groups have similar preferences, but better when their preferences differ more, e.g., Figure \ref{fig:eps_comp}. In theory, the neural classifier should be equipped to learn the same predictor since it is given the Demographic Ratios as input, but in practice, we see that this fails. This was further supported by toy examples where we explicitly trained classifiers in scenarios where taking the mean of the ratio features would result in AUC of about 0.6, and where we also added noisy dimensions for each \textit{item} to simulate the item embeddings it could ignore to match Demographic Ratio AUC. Note that one typically does not train classifiers in such settings since the best possible classifiers would only perform slightly better than random choices. The neural models were very sensitive to the added noise and routinely failed to achieve AUC greater than 0.5 for recommendation/item lists of 10 or more items and a few dimensions of noise for each item. 

While we believe it would be challenging to design and train neural classifiers that would outperform both of our approaches across all/smaller Representation AUC, we cannot exclude the possibility that an ideal classifier may be able to match Recommendation AUC with the Representation AUC for some models. However, it is worth noting that a classifier may pick up minor differences in the representation space despite not affecting the output rankings. In theory, representations that produce two identical sets of recommendations may be fully separable, e.g., least significant bit encoding. Figures \ref{fig:auc_synth1}, \ref{fig:auc_real2}, \ref{fig:auc_synth2} etc, all reveal that the same Representation AUC can be associated with different Recommendation AUC for different models, i.e., the same Representation AUC can lead to various levels of demographic differences in the output. This implies that not all the demographic knowledge in the representation is propagated to the output for at least one of the models. Finally, since we are looking at a limited part of the output (top 40 recommendations), unmeasured demographic differences may be found in the excluded part of the output.

\subsection{Which Recommendation/Representation AUC Measure To Trust?}
We have shown that different approaches for calculating Recommendation AUC can produce different values. We argue that one should trust the method that produced the best (average) AUC value, regardless of the model complexity. In practice, we often see that neural models can overfit and return a lower AUC value than Logistic Regression or Demographic Ratio AUC. The latter two methods cannot capture correlations that are not linearly separable, but this is of little importance if they outperform models that are.

Despite this conclusion, we have presented two different measures of Recommendation AUC throughout this paper. This is because we want to properly explore how the two proposed measures differ for various models and datasets, and to highlight further the challenges of accurately measuring Recommendation AUC. If we aggregated a single value through max, it would be more difficult to analyze why models or datasets result in specific trends in Recommendation AUC.

\subsection{Optimal Representation AUC in other Research?}
Some existing research achieved optimal AUC for datasets like Movielens, while our best model achieved around 0.6 AUC for both Gender and Age. Our experience is that all of these claims have the following explanations:
\begin{enumerate}
    \item In most evaluation scenarios, the model will only recommend for users present in the training data. These models can perfectly learn all user-demographic correlations and filter these out. This is a valid strategy for these evaluation scenarios, but it does not generalize to new users and will remove all spurious correlations of demographics and users. Known users' demographics are perfectly filtered, so the model will achieve AUC = 0.5 for these users.
    \item If the classifier used to measure AUC is severely overfitted to the training data, it may treat unseen representations randomly, which results in an AUC of 0.5 for these users.
\end{enumerate}
In our evaluation setup, we exclusively test on new users not present in the training set, so all aspects of the evaluated models must generalize to new input.

\section{Conclusion}
Our results show that optimising for neutral representations makes the treatment of different demographic groups more similar, but raises questions about evaluating these methods by Representation AUC. Firstly, in both real and synthetic datasets with demographic differences, we have shown that the Recommendation AUC trend towards perfect while Representation AUC still indicate that the Representations encode demographics. At perfect Recommendation AUC, further optimization of Representation AUC will not positively affect recommendation parity, and the likely worsening of recommendation performance will be for nothing. Secondly, two models achieving the same Representation AUC have been shown to achieve substantially different Item ratio, Kendall-Tau and Recommendation AUC. We have also observed that fair models optimized for datasets with different levels of demographic information will achieve the same Recommendation AUCs as non-fair models, but for larger Representation AUCs. Finally, we have highlighted some issues with the AUC metrics, namely, how they easily overfit, how the same value can have multiple explanations, and how different approaches for calculating this metric are more accurate for different levels of demographic information. Even if all of these issues are handled in the best way possible, our findings indicate that Representation AUC at best serves as an upper limit of how much demographics can influence the recommendations, without capturing that different models with the same upper limit can express different levels of demographic information in their recommendations. Thus, we suggest limiting the usage of the Representation AUC when evaluating models optimized for neutral representations. If it is used, it should be used together with metrics that also compare the model's output.

Our results also shed light on how fair heuristics like Popularity and Random baselines are deemed fair and discriminatory by different fairness metrics. \texttt{RAND} performs very well on Recommendation AUC, good on Item ratio, but achieves the worst possible Kendall-Tau score. \texttt{POP} performs very well on Item ratio and Kendall-Tau, but not on Recommendation AUC. Further, certain experiments illustrate an issue with evaluating such heuristics in recommender systems that do not allow repeated recommendations, since the heuristic becomes conditioned on the known user data. The conditioned heuristics will not behave fairly unless the known data is unbiased by demographics. If an item is more popular among one group, a method poised to recommend it equally for all groups may recommend it more often to the other group in which fewer users have interacted with it in the training data. While these observations are only made for these simple baselines, they are equally valid for a (theoretical) perfectly fair recommender system that completely disregards demographics. A perfectly fair heuristic will not achieve the perfect metric score for all of our considered metrics since the (input) data-conditional recommendations will correlate with demographics whenever the input data does. How models behave for different metrics as they close in on being perfectly fair remains an open question. For example, will the item ratio move towards zero before increasing again as it gets fairer, or will it monotonically decrease without visiting zero? Answering these questions will be pivotal should perfectly fair recommender systems ever be proposed.

Finally, we have explored metrics for evaluating how different the recommendations given to different demographic groups are and how they behave for different datasets and model parameters. We have proposed two methods for measuring Recommendation AUC and thoroughly compared them and how they relate to Representation AUC. Our extensive experiments highlight the strengths and weaknesses of existing and new metrics.

\section*{Acknowledgments}
This publication has been partly funded by the SFI NorwAI, (Centre for Research-based Innovation, 309834). The authors gratefully acknowledge the financial support from the Research Council of Norway and the partners of the SFI NorwAI.


\appendix
\section{Additional Results}

\begin{figure}
    \centering
    \includegraphics[width=1.0\linewidth]{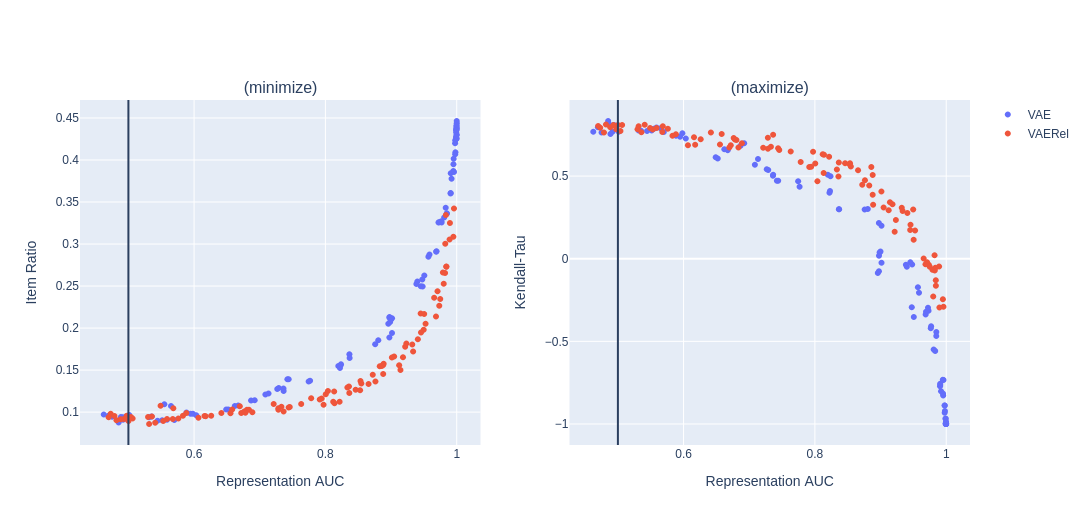}
    \caption{Item Ratio and Kendall-Tau plotted against Representation AUC. Comparison of \texttt{VAE} and \texttt{VAERel} on synthetic datasets with different $\epsilon$ parameters.}
    \label{fig:item_kendall_comp}
\end{figure}
\begin{figure}
    \centering
    \includegraphics[width=1.0\linewidth]{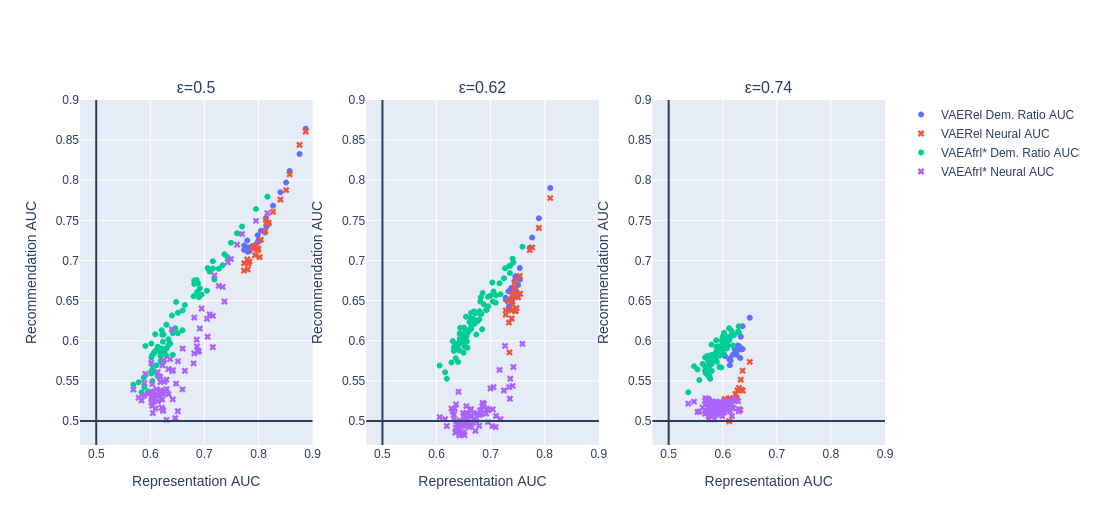}
    \caption{Recommendation AUC plotted against Representation AUC in synthetic datasets with three different $\epsilon$ parameters.}
    \label{fig:recauc_multi_eps}
\end{figure}

Figure \ref{fig:item_kendall_comp} reinforces that the Representation AUC is not a good proxy for recommendation fairness since different models achieve the same Representation AUC for different Item ratios and Kendall-Tau values. The plot also shows how both metrics have an exponential relationship with Representation AUC.

We provide three plots similar to Figure \ref{fig:eps}, where we compare both measures of Recommendation AUC of \texttt{VAE} and \texttt{VAERel} for three different synthetic datasets. The plots have the same scale and decreasing levels of demographic preferences overlap.


\bibliographystyle{unsrt}  
\bibliography{references}

\end{document}